\documentclass[sigconf]{acmart}
\usepackage{booktabs}
\usepackage{tabularx}
\usepackage{tabto}
\usepackage[T1]{fontenc}
\usepackage{subcaption}
\usepackage{graphicx}
\usepackage{hyperref}
\usepackage{color}

\setcopyright{none}
\renewcommand\footnotetextcopyrightpermission[1]{} 

\begin{document}

\title{Creating Structure in Web Archives With Collections: Different Concepts From Web Archivists}

\author{Himarsha R. Jayanetti}
\email{hjaya002@odu.edu}
\orcid{0000-0003-4748-9176}
\affiliation{%
  \institution{Old Dominion University}
  \streetaddress{5115 Hampton Blvd}
  \city{Norfolk}
  \state{Virginia}
  \country{USA}
  \postcode{23529}
}

\author{Shawn M. Jones}
\email{smjones@lanl.gov}
\orcid{0000-0003-4748-9176}
\affiliation{%
  \institution{Los Alamos National Laboratory}
  \state{New Mexico}
  \country{USA}
  \postcode{87545}}
  
\author{Martin Klein}
\email{mklein@lanl.gov}
\orcid{0000-0003-0130-2097}
\affiliation{%
  \institution{Los Alamos National Laboratory}
  \state{New Mexico}
  \country{USA}
  \postcode{87545}}
  
\author{Alex Osbourne}
\email{aosborne@nla.gov.au}
\affiliation{%
  \institution{National Library of Australia}
  \city{Canberra}
  \state{ACT}
  \country{Australia}}

\author{Paul Koerbin}
\email{pkoerbin@nla.gov.au}
\affiliation{%
  \institution{National Library of Australia}
  \city{Canberra}
  \state{ACT}
  \country{Australia}}

\author{Michael L. Nelson}
\email{mln@cs.odu.edu}
\orcid{0000-0003-3749-8116}
\affiliation{%
  \institution{Old Dominion University}
  \streetaddress{5115 Hampton Blvd}
  \city{Norfolk}
  \state{Virginia}
  \country{USA}
  \postcode{23529}
}

\author{Michele C. Weigle}
\email{mweigle@cs.odu.edu}
\orcid{0000-0002-2787-7166}
\affiliation{%
  \institution{Old Dominion University}
  \streetaddress{5115 Hampton Blvd}
  \city{Norfolk}
  \state{Virginia}
  \country{USA}
  \postcode{23529}
}

\begin{abstract}
As web archives' holdings grow, archivists subdivide them into collections so they are easier to understand and manage. In this work, we review the collection structures of eight web archive platforms: : Archive-It, Conifer, the Croatian Web Archive (HAW), the Internet Archive's user account web archives, Library of Congress (LC), PANDORA, Trove, and the UK Web Archive (UKWA). We note a plethora of different approaches to web archive collection structures. Some web archive collections support sub-collections and some permit embargoes. Curatorial decisions may be attributed to a single organization or many. Archived web pages are known by many names: mementos, copies, captures, or snapshots. Some platforms restrict a memento to a single collection and others allow mementos to cross collections. Knowledge of collection structures has implications for many different applications and users. Visitors will need to understand how to navigate collections. Future archivists will need to understand what options are available for designing collections. Platform designers need it to know what possibilities exist. The developers of tools that consume collections need to understand collection structures so they can meet the needs of their users.
\end{abstract}


\maketitle
\pagestyle{plain}

\section{Introduction}
Researchers, including journalists \cite{mask_stockpile_ars_2020,buffalo_shooter_huffpo_2022}, social scientists \cite{arms2006,MEET:MEET14504801096}, and historians \cite{milligan_history_2019} increasingly make use of web archives. Web archives preserve the content of web pages as they were at a specific point in time as \textbf{mementos}. Web archives are vast, the largest consisting of billions of documents \cite{kahle_wayback_count_2020}. Collections are a common organizational technique employed to bring order to this vastness. Web archive collections consist of web pages that were hand-picked by archivists and subject matter experts to represent a specific topic. Using collections simplifies management for archivists and allows them to showcase content, making patrons aware of the collection as well as their web archiving organization as a whole. Patrons benefit from collections because they have an intelligently selected set of mementos to review that supports their topic of interest. 

We recognize that the term \emph{collection} has many definitions. In the scope of this paper, we define \textbf{collection} based on a web archive platform's front-end presentation of a set of mementos that are grouped by topic. Figure \ref{fig:archive-it-collection} shows a screenshot of a collection landing page for Archive-It's collection \emph{Environmental Justice}\footnote{\url{https://archive-it.org/collections/7635}}. This page contains a list of resources that a visitor can examine, along with metadata about the collection. Figure \ref{fig:pandora-collection} contains a screenshot of the collection landing page of PANDORA's \emph{Indigenous Australians} collection\footnote{\url{https://pandora.nla.gov.au/subject/12}}.  At PANDORA, a visitor can view the list of page titles, but also how this collection fits within an overall hierarchy of topics, sub-collections, and subcategories. Similarly Figures \ref{fig:haw-collection}, \ref{fig:ukwa-collection}, \ref{fig:ia-collection}, and \ref{fig:conifer-collection} show collection landing pages\footnote{\url{https://haw.nsk.hr/en/category/biology/}}\footnote{\url{https://www.webarchive.org.uk/en/ukwa/collection/2387}} from the Croatian Web Archive (HAW) the United Kingdom Web Archive (UKWA), an Internet Archive (IA) user account\footnote{\url{https://archive.org/details/@shawnmjones?tab=web-archive}}, and Conifer\footnote{\url{https://conifer.rhizome.org/despens/usa-today-the-wall}} respectively. We differentiate collections from the \textbf{greater web archive} -- the web archiving platform as a whole. For example, PANDORA would be the greater web archive containing \emph{Indigenous Australians}.

\begin{figure*}
\centering
    \begin{subfigure}{0.5\hsize}
    \centering
    \includegraphics[width=\textwidth]{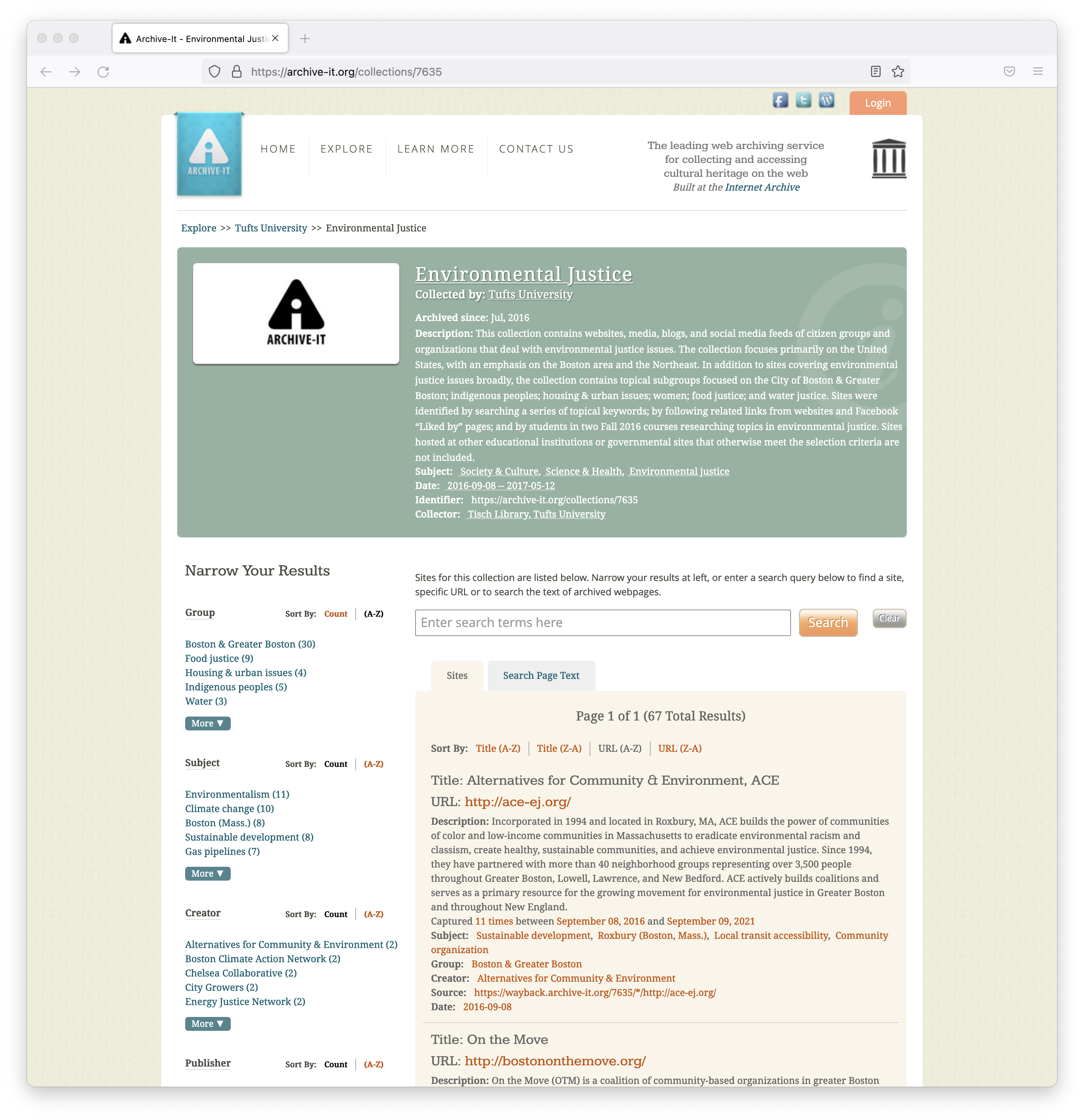}
    \caption{Archive-It's \emph{Environmental Justice}}
    \label{fig:archive-it-collection}
    \end{subfigure}%
    
    \begin{subfigure}{0.5\hsize}
    \includegraphics[width=\textwidth]{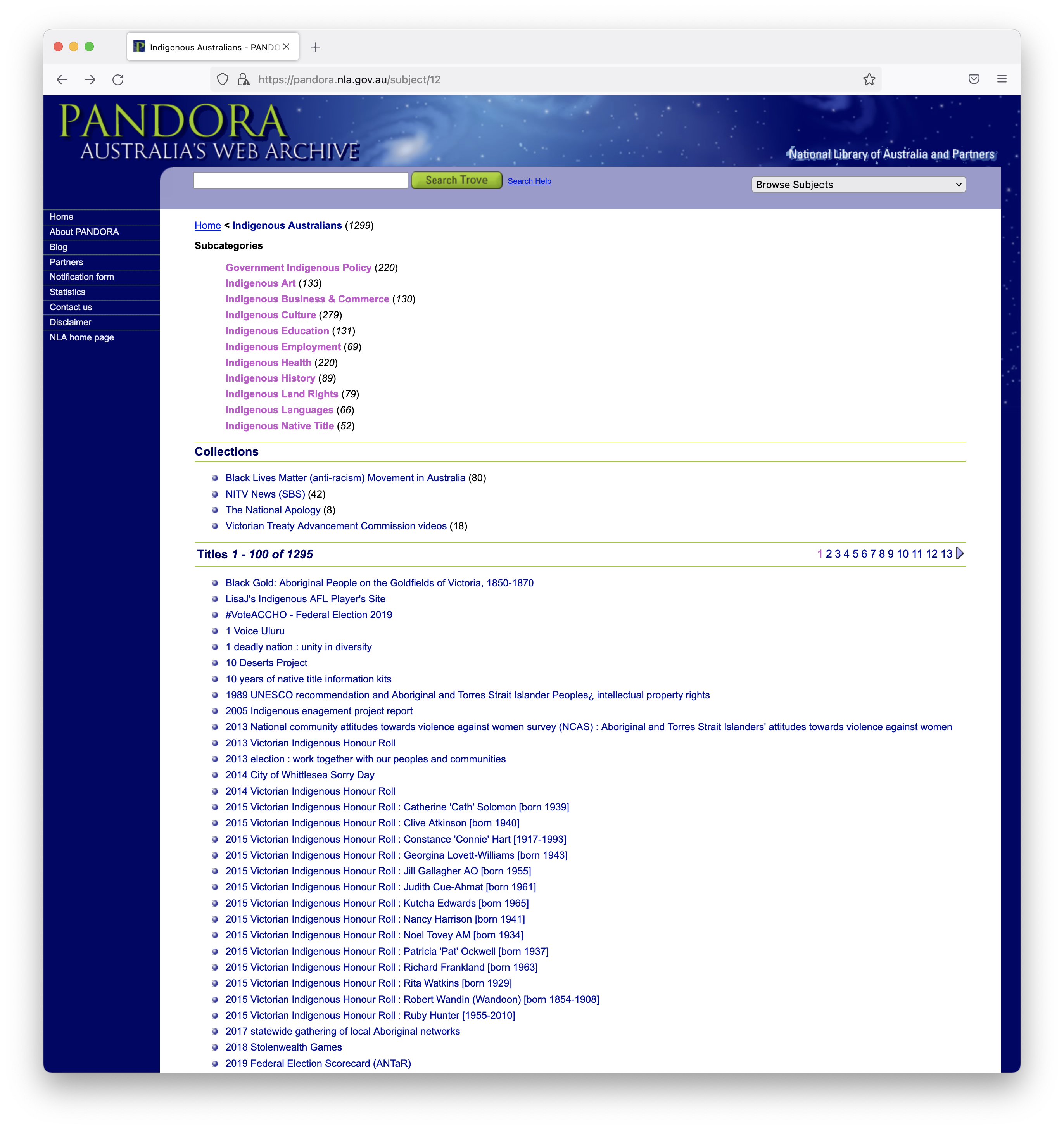}
    \caption{PANDORA's \emph{Indigenous Australians}}
    \label{fig:pandora-collection}
    \end{subfigure}
\caption{Screenshots of some example web archive collection landing pages.}
\end{figure*}

\begin{figure*}\ContinuedFloat
\centering
    \begin{subfigure}{0.5\hsize}
    \centering
    \includegraphics[width=\textwidth]{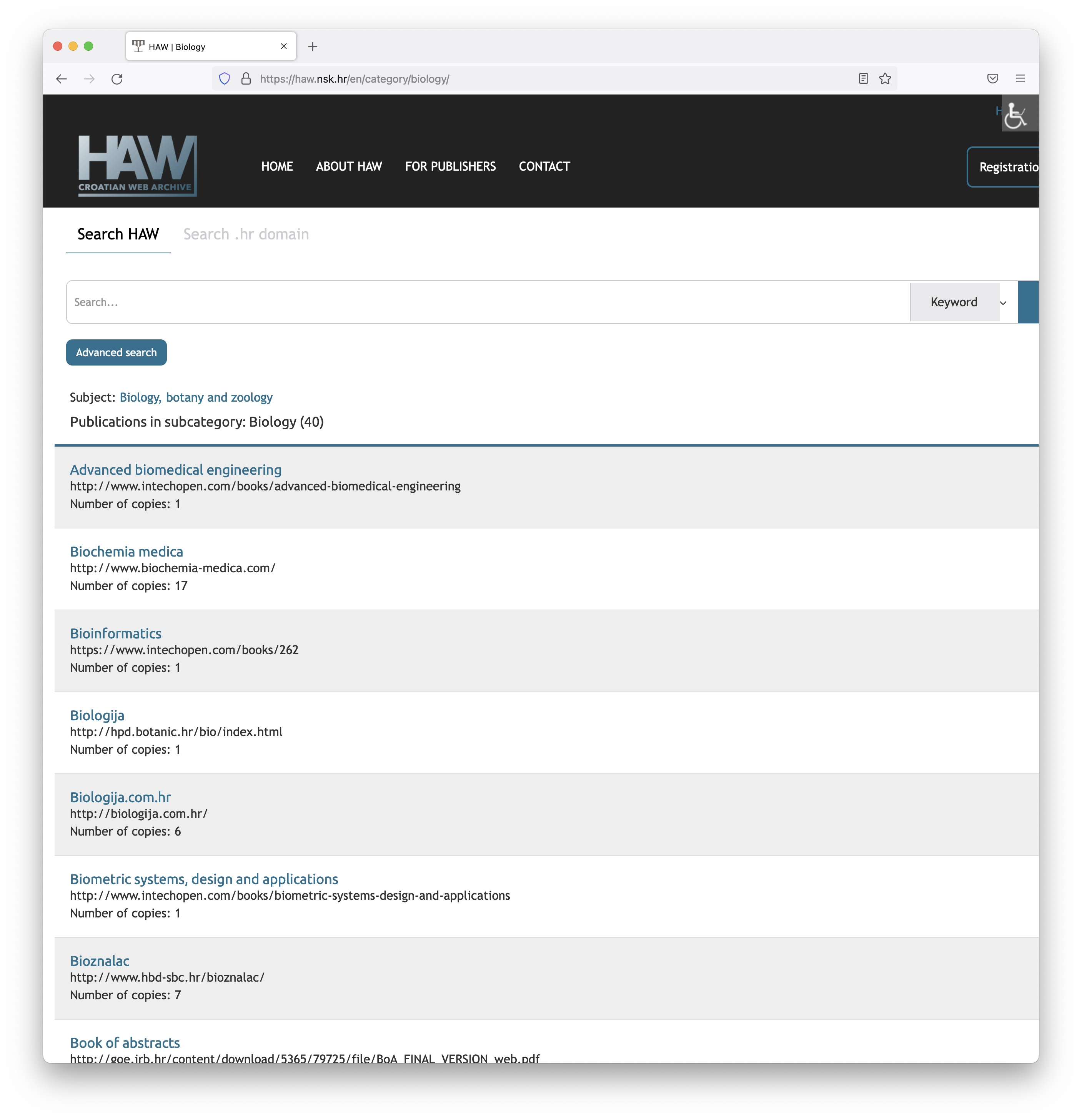}
    \caption{HAW's \emph{Biology}}
    \label{fig:haw-collection}
    \end{subfigure}%
    
    \begin{subfigure}{0.5\hsize}
    \includegraphics[width=\textwidth]{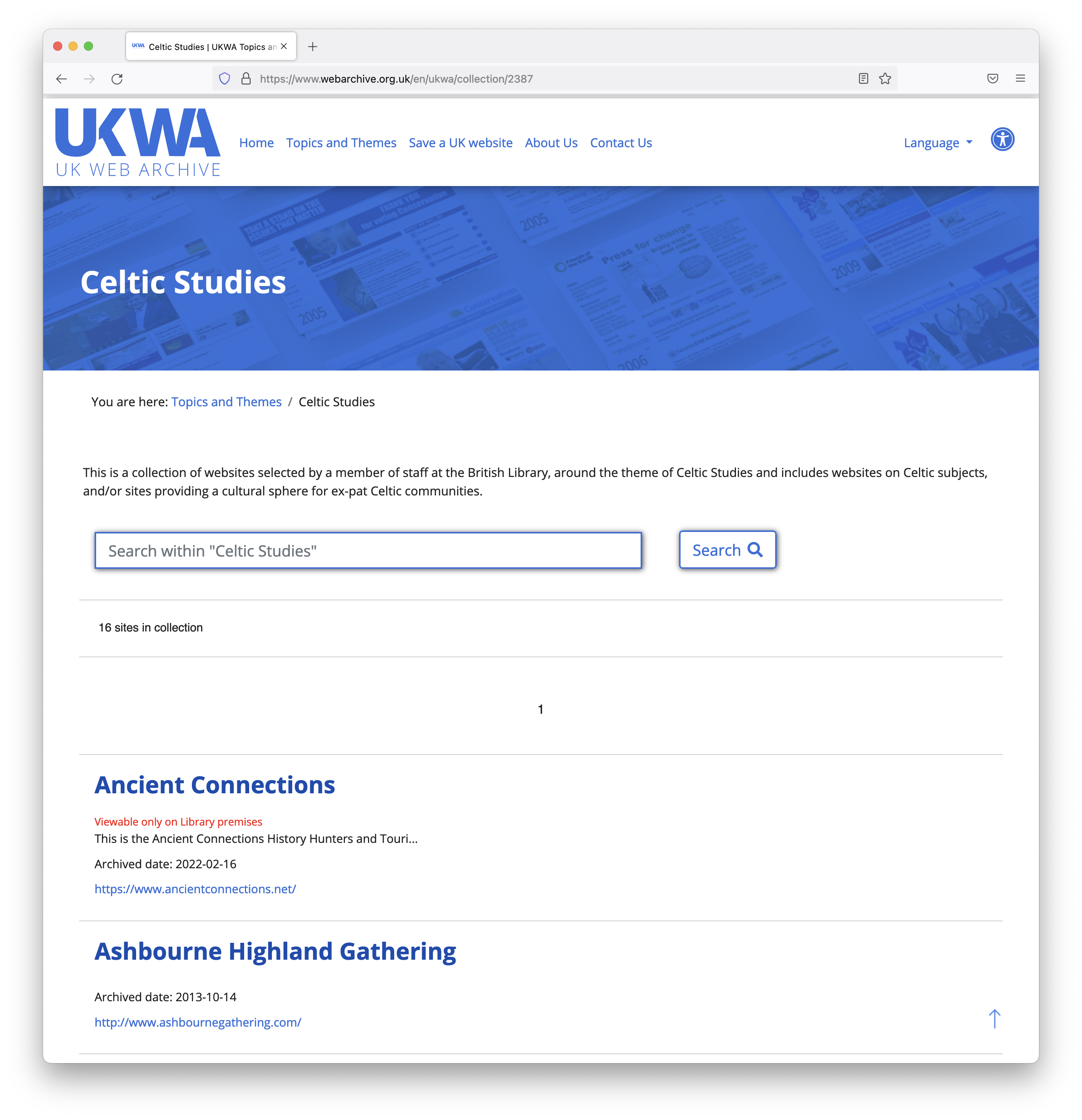}
    \caption{UKWA's \emph{Celtic Studies}}
    \label{fig:ukwa-collection}
    \end{subfigure}
\caption{Screenshots of some example web archive collection landing pages.}
\end{figure*}

\begin{figure*}\ContinuedFloat
\centering
    \begin{subfigure}{0.5\hsize}
    \centering
    \includegraphics[width=\textwidth]{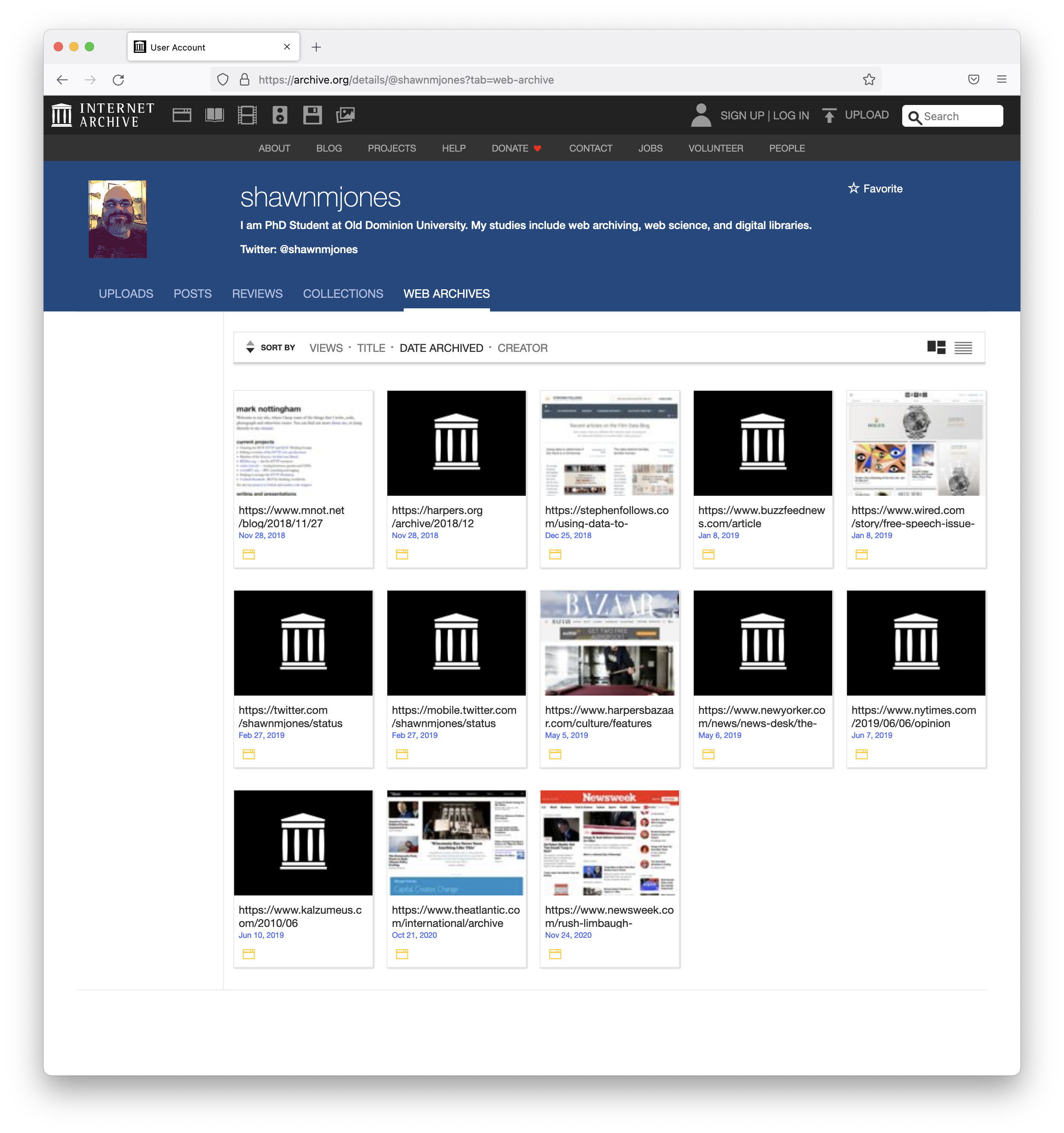}
    \caption{IA's user account web archives for one of this paper's authors}
    \label{fig:ia-collection}
    \end{subfigure}%
    
    \begin{subfigure}{0.5\hsize}
    \includegraphics[width=\textwidth]{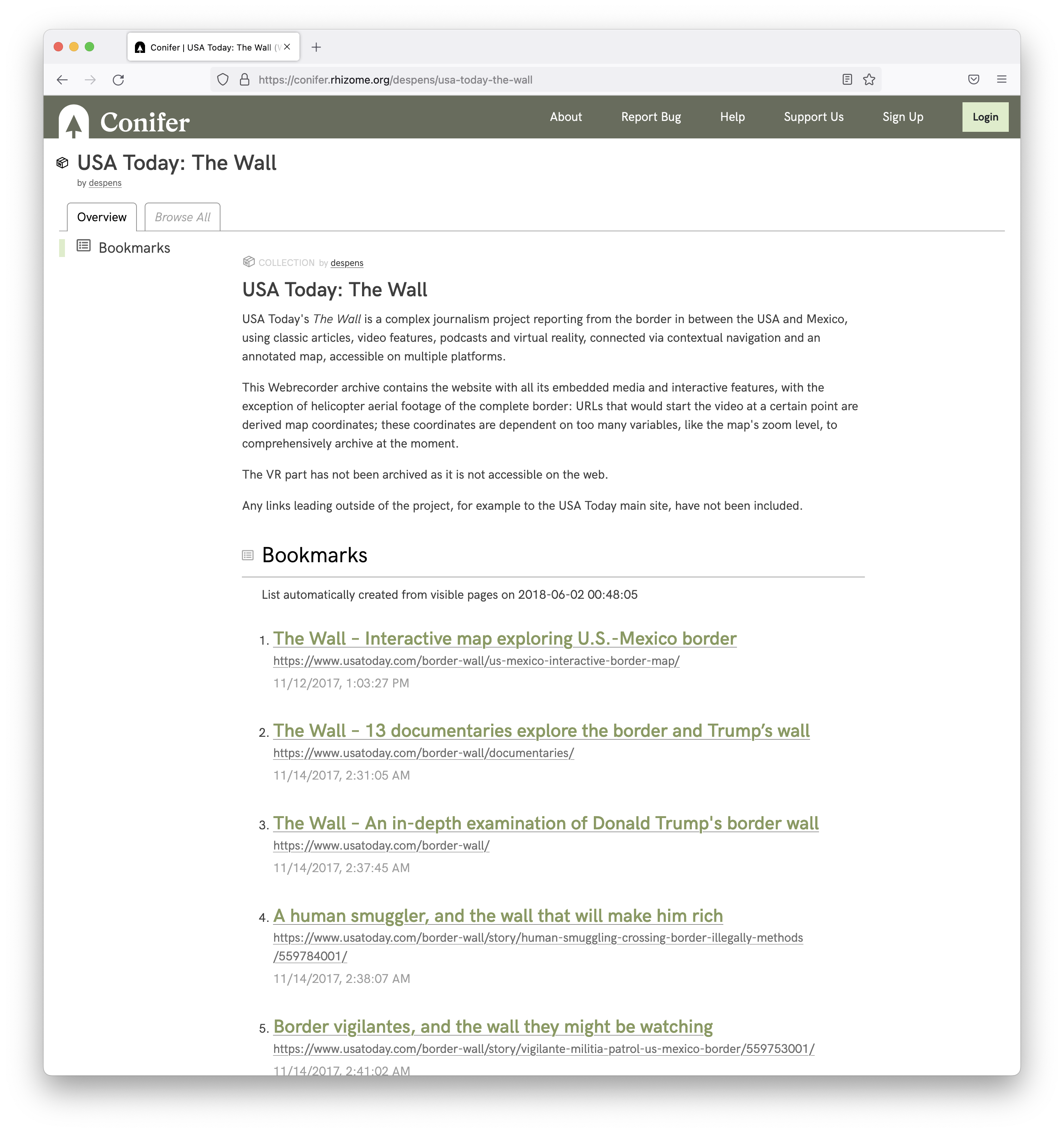}
    \caption{Conifer's \emph{USA Today: The Wall}}
    \label{fig:conifer-collection}
    \end{subfigure}
\caption{Screenshots of some example web archive collection landing pages.}
\label{fig:landing_pages}
\end{figure*}

Even in the examples shown in Figure \ref{fig:landing_pages} there are differences in approaches to these collections. If a memento is created in an Archive-It collection, it is not shared between collections, whereas, with PANDORA, this is possible. Design decisions like this represent different \textbf{collection structures} that serve as models for how an archivist or platform designer might thematically organize their mementos. Collection structures have implications for how visitors interact with the collection. For example, does a visitor first visit the landing page to view a set of page titles and then decide among mementos for that page, or does the collection directly present links to mementos? Additionally, collection structures present different challenges to authors of third-party tools that consume and analyze these collections as part of Big Data efforts.


While the collection structure of each platform may vary depending on their organization's requirements and design choices, each structure shares some basic elements. In this work, we define and standardize nomenclature so we can discuss these elements. Our goal is not to prescribe how a collection should be designed, but rather to understand the different collection structures already in existence. Thus, we reviewed the collection structures of Archive-It\footnote{\url{https://archive-it.org/}}, the National Library of Australia's (NLA) PANDORA\footnote{\url{http://pandora.nla.gov.au/}} and Trove\footnote{\url{https://trove.nla.gov.au/}} archives, the Croatian Web Archive (HAW)\footnote{\url{https://haw.nsk.hr/}}, the Library of Congress Web Archive (LC), the United Kingdom Web Archive (UKWA)\footnote{\url{https://www.webarchive.org.uk/en/ukwa/index}}, Conifer\footnote{\url{https://conifer.rhizome.org/}} (formerly Webrecorder\cite{webrecorder_replay_2020}). Finally, we include the Internet Archive's\footnote{\url{https://archive.org}} (IA) user account web archives because IA's \emph{Wayback Machine} is synonymous with web archiving, even though its collections are tied to a specific user rather than a theme. While there are many web archiving initiatives \cite{gomes_2011,enwiki:1080525996}, we focused on these eight platforms because they provide collections as defined above. Through this review, we address the following research questions:

\begin{itemize}
  \item RQ1: What different collection structures exist?
  \item RQ2: What do these distinct collection structure approaches have in common?
\end{itemize}

Existing work has focused on the nature of digital collections \cite{fenlon_toward_2017}, user behavior when curating personal collections \cite{MULL2014192,doi:10.1177/2056305116662173}, the behavior of archivists \cite{ogden2017observing}, the capabilities of web archive platforms \cite{niu2012}, and the challenges of building collections with Archive-It \cite{doi:10.1086/669993,doi:10.1086/685975}. Our work augments this by focusing on existing web archive collection structures for the benefit of future archivists as well as web archive platform and analytics tool designers. We recognize that many organizations create collections with Archive-It, but we produced this work to highlight the novel concepts of many different platforms, including those from national libraries like the Library of Congress (LC), the National Library of Australia (NLA) (through PANDORA/Trove), and the British Library (through the UKWA). Our contributions are as follows:
\begin{itemize}
\item Documentation of the collection structures followed by different web archives, not only to help current archivists and developers understand the present state, but also to consolidate and summarize knowledge for platform developers.
\item An analysis of the similarities among these various web archive collection structure approaches to provide ideas to future platform developers.
\end{itemize}

This work is provided to assist collection analysis software
projects like the Off-Topic Memento Toolkit \cite{alnoamany_detecting_2016,jones_off-topic_2018_nonanon}, Hypercane \cite{jones2021hypercane,jones2021hypercane2}, and ArchiveSpark \cite{holzmann_archivespark:_2016}.

\section{Background}

\begin{figure*}
\centering
    \begin{subfigure}{0.5\hsize}
    \centering
    \includegraphics[width=\textwidth]{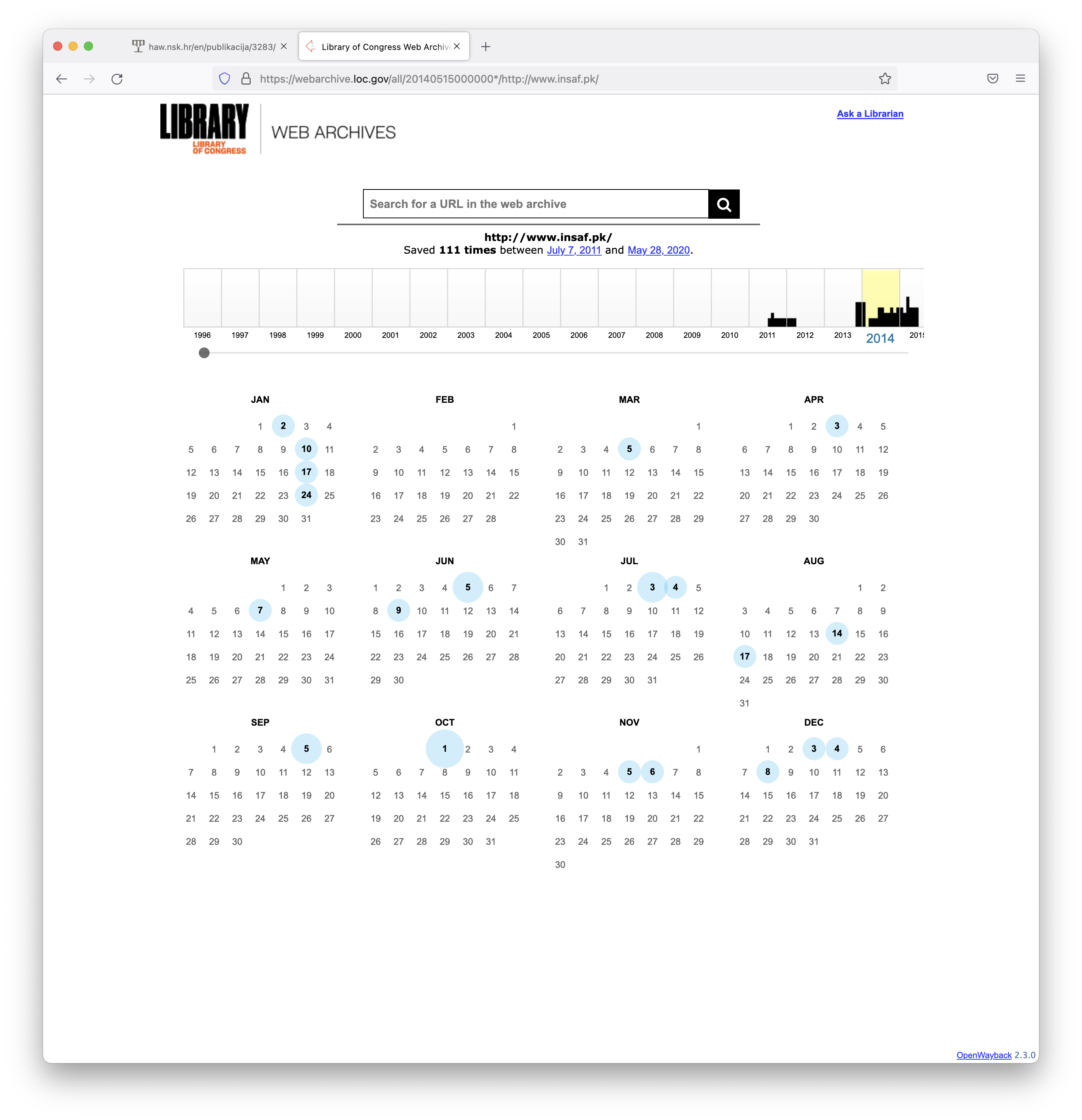}
    \caption{Library of Congress (LC)}
    \label{fig:haw-human-readable-timemap}
    \end{subfigure}%
    
    \begin{subfigure}{0.5\hsize}
    \includegraphics[width=\textwidth]{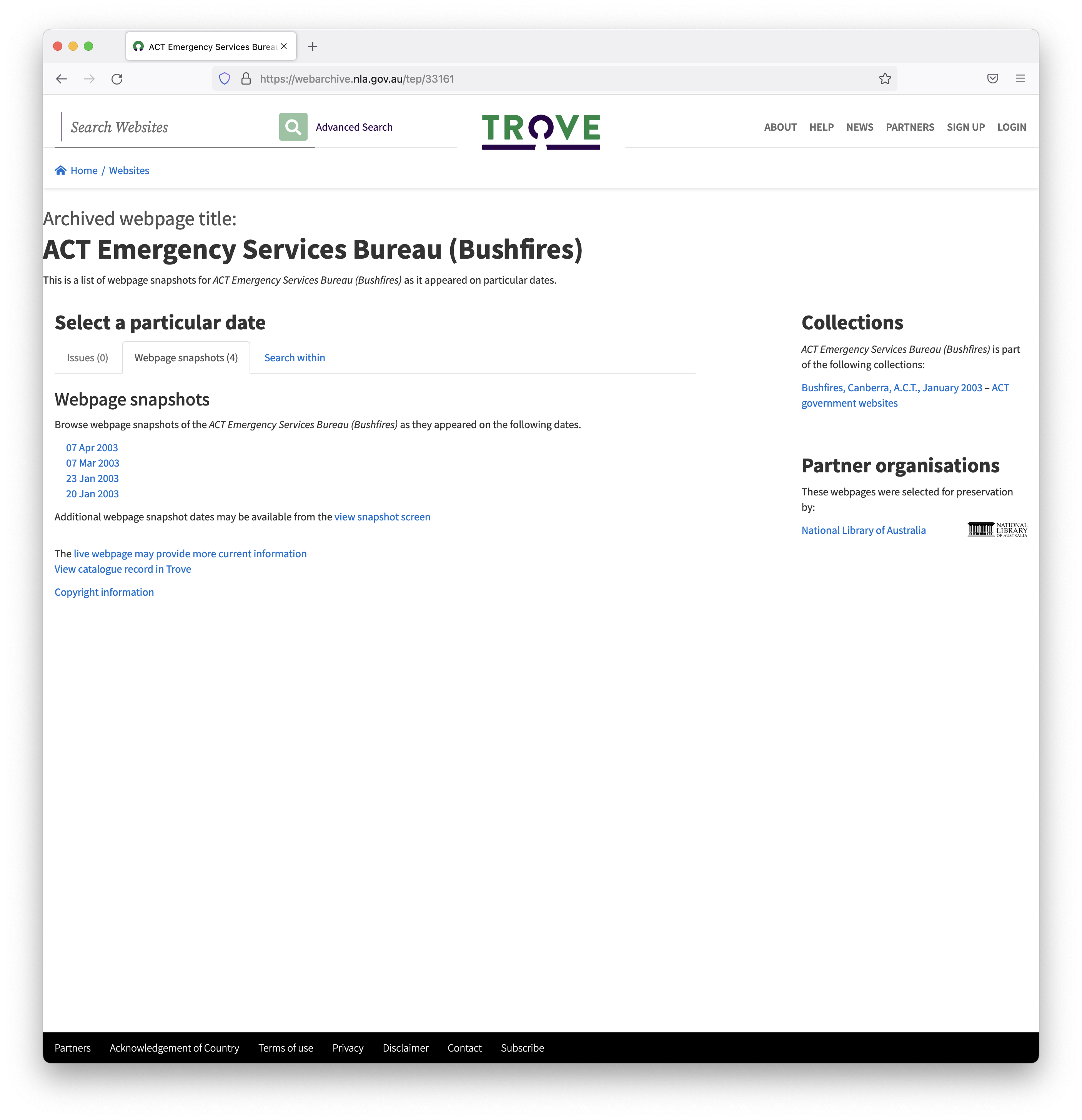}
    \caption{Trove Title Entry Page (TEP)}
    \label{fig:tep-human-readable-timemap}
    \end{subfigure}
\caption{Screenshots showing examples of human-readable TimeMaps.}
\label{fig:human-readable-timemaps}
\end{figure*}

When building a web archive collection, an archivist selects a set of URIs as seeds.  Each seed is an \textbf{original resource} that reflects the current state of the web resource. Each memento is an observation of that resource at a particular point in time, its \textbf{memento-datetime}. Each original resource is identified by its URI-R (e.g., \url{https://www.cnn.com}) and each of its mementos is identified by a URI-M (e.g., \url{https://wayback.archive-it.org/7678/20190319204514/https://www.cnn.com/}). A \textbf{TimeMap} is a listing of the mementos created for an original resource, including the URI-M of each memento and its memento-datetime. \textbf{Human-readable} TimeMaps are rendered as a list or calendar with links to each URI-M. Some examples of human-readable TimeMaps are shown in Figure \ref{fig:human-readable-timemaps}. \textbf{Machine-readable} TimeMaps can take a variety of formats, such as JSON. Many web archives are compliant with the Memento Protocol \cite{van_de_sompel_rfc_2013}, which formalizes these concepts and provides standardized methods of linking mementos, original resources, and TimeMaps.

Not all original resources are seeds. An archivist can instruct the web archiving platform to follow links from a seed to other original resources and capture those as well. \textbf{Seed mementos} are mementos that the archivist directly asked the platform to capture. \textbf{Deep mementos} are mementos captured by crawling a seed memento's links. We make this distinction because a visitor can immediately discover the seed mementos through a web archive collection's user interface, but may need to click links from seed mementos to discover deep mementos. Tools attempting to capture information about a collection are also limited by this distinction because seed mementos are advertised through the user interface while deep mementos must be crawled to be discovered. Archive-It is an example of a platform that requires this distinction.

\section{Related Work}
Much work exists in analyzing the use and creation of collections. Fenlon \cite{fenlon_toward_2017} performed an in-depth study of two digital collections, detailing different approaches to data models, supporting context, and overall visualization of content. She notes that all collection curators may learn from the structures of existing digital collections. She details how collections have implications for scholarly communications and that the goal of a collection influences its collection structure. Though Fenlon did not analyze web archives, her work has inspired our technical analysis of collection structures.

Mull and Lee \cite{MULL2014192} applied the Users and Gratifications model \cite{doi:10.1111/j.00117315.2004.02524.x} to understand why Pinterest users select certain items for their collections. They contrasted the behavior of Pinterest users with other social media platforms and found that image-sharing platforms have unique factors. Wang et al. \cite{doi:10.1177/2056305116662173} completed a similar study to understand how users not only created collections, but interacted with them on Pinterest. Ogden et al. \cite{ogden2017observing} studied web archivists themselves to better understand the ways in which they ``shape and maintain the preserved Web.'' Nwala et al. \cite{nwala2018} analyzed how to leverage search engine results to populate web archive collections. Nwala et al. \cite{nwala_bootstrapping_2018} also used Archive-It collections to compare human-made vs. automatically or semi-automatically generated collections. Klein et al. investigated the possibility of performing focused crawls to build collections from greater web archives \cite{10.1145/3201064.3201085}. Where those studies focused on creating and curating collections, or understanding archivists' motivations for doing so, our work analyzes what is present already and the behavior of platform designers as revealed through collection structures.

Web archives and the challenges they face have been extensively examined in the past. Crook \cite{doi:10.1108/02640470910998542} detailed the state of web archiving in Australia in 2009, noting the challenges with establishing different capture efforts as part of the PANDORA archive. Slania \cite{doi:10.1086/669993} and Deutch \cite{doi:10.1086/685975} detailed their experiences with using Archive-It to archive art web sites. In 2012, Niu \cite{niu2012} conducted an analysis of 10 different web archive platforms and discovered that search capabilities by URL and keyword were common, with varying levels of capability, but none provided data mining services at that time. Our work is similar in that we are analyzing the capabilities of different web archives, but, unlike the work of others, we are focusing on the subset of web archives that offer themed collections, and we provide a model for understanding their different collection structures.


How users directly leverage web archive collections has received attention. Milligan \cite{milligan2016lost} discussed how historians of the present and future might benefit from understanding large collections in the Internet Archive (IA) and Archive-It. According to Risse \cite{risse2014you} and Gossen et al. \cite{gossen2016analyzing}, scientists are generally interested in examining smaller and more targeted event-centric collections of documents available in a web archive. They discuss the difficulties of working with web archives and provide a research methodology for extracting and analyzing archive sub-collections focusing on specific subjects and events. Jones et al. \cite{jones2018many} focused on web archive collections' structural characteristics to better comprehend them. They applied concepts from AlSum's work \cite{alsum_thumbnail_2014} to demonstrate how the growth of collections could be compared by quantifying when mementos and seeds were added, including their age and frequency of creation. This work is similar because we are analyzing the structures within web archives that support collections, but we differ in several ways. Milligan, Risse, and Gossen discussed how visitors would consume collections, but did not analyze their structures. Jones analyzed the structural features but not the structures themselves.

Ke et al. \cite{ke2008toward} semantically leveraged collection structures to aid user exploration of massive data corpora through clustering. They developed the LAIR2 clustering method and the prototype LAIR2 Scatter/Gather \cite{10.1145/133160.133214} browser for exploring collections. Through analyzing user behavior with the scatter/gather concept and health information searches, Zhan et al. \cite{zhang2014evaluation} learned that users' mental models of search have a significant influence in how they utilize search interfaces. Our work exists to support such efforts by detailing other models of collection presentation. 

Other research examined the fundamental properties of web archive collections. Padia et al. \cite{padia_visualizing_2012} created several visualization techniques to help users better grasp collection characteristics. AlNoamany et al.  \cite{alnoamany2017generating} pioneered the concept of combining social media stories with web archive collections to provide a user-friendly interface for corpus summarization. She asked domain experts to manually select mementos that represented a collection. She then developed an algorithm that took into account the collection structure to automatically select mementos. Test subjects could not tell the difference between stories generated by her algorithm or those generated by domain experts. Our work analyzes the collection structures that make this type of visualization and summarization possible.

Hypercane \cite{jones2021hypercane,jones2021hypercane2} is a toolkit that uses intelligent sampling to summarize large web archive collections. This software uses the structural aspects of the collection and the content of the collection's mementos to automatically select mementos that are representative of a web archive collection. Hypercane relies on the AIU \cite{jones2018aiu_nonanon} library to discover seeds within Archive-It, PANDORA, and Trove collections. All of these web archive platforms have adopted Memento \cite{van_de_sompel_rfc_2013}, but they have no standard method for clients to discover seeds and metadata. AIU accepts a collection identifier and then scrapes the associated collection landing page for that data. The Off-Topic Memento Toolkit (OTMT) \cite{jones_off-topic_2018_nonanon} applies textual similarity metrics to determine which mementos in a collection have gone off-topic (e.g., crawled as 404s, database error pages, no longer matching the collection topic). The OTMT analyzes the mementos within a given TimeMap, but needs to know which TimeMaps belong to a web archive collection, hence it also relies on AIU. Jones et al. also investigated how users might better understand web archive collections by using social cards \cite{jones2019social}. They discuss how surrogates representing a subset of a collection can help users determine what information needs a collection might satisfy. From this work they developed Raintale \cite{jones_mementoembed_raintale_2020} that creates surrogates for groups of mementos. Combining Hypercane's sampling with Raintale's visualization capability provides summaries of web archive collection in formats easily understood by web users. These tools all belong to the Dark and Stormy Archives (DSA) project \cite{jones_dissertation_nonanon,dsa_code4lib_2022} which focuses on analyzing and summarizing web archive collections. Our work exists to help efforts like the DSA that consume and analyze web archive collections.




\section{Web Archive Collection Structures}

In December 2021, we chose eight web archives to analyze because they support our definition of a collection. We looked at the web archive platforms at Archive-It, Conifer, the Croatian Web Archive (HAW), the Internet Archive's user account web archives, Library of Congress (LC), the National Library of Australia (NLA: PANDORA and Trove web archive platforms), and the UK Web Archive (UKWA). 

Table~\ref{tab:collection_structures_per_archive_platform} shows the collection structures of the different platforms we analyzed, addressing RQ1. Here, we have a summary of the many behaviors that represent the requirements of various web archiving platforms discussed earlier.

\begin{table*}[t]
\caption{Details on different web archive platform collection structures.}
\label{tab:collection_structures_per_archive_platform}
\centering
\resizebox{\textwidth}{!}{%
\begin{tabular}{>{\raggedright}p{2cm} >{\raggedright}p{2cm} >{\raggedright}p{2cm} >{\raggedright}p{2.5cm} >{\raggedright}p{2cm} >{\raggedright}p{2cm} >{\raggedright}p{2cm} >{\raggedright}p{2cm} p{2cm} p{2cm} }
\toprule
\textbf{Collection Platform} & \textbf{Name for mementos}          & \textbf{Sub-collections?} & \textbf{Attribution}         & \textbf{Private collections supported?} & \textbf{Mementos are accessible from more than one collection?} & \textbf{Deep mementos accessible from within the collection?} & \textbf{Human-readable TimeMap membership} & \textbf{Embargoed resources?} & \textbf{Navigational hierarchy} \\ 
\midrule
Archive-It                   & Captures                            & No                        & Single account               & Yes                                     & No*                                                             & Yes                                                           & Collection                  & No                            & Type 1\\\midrule
Conifer                      & Captures                            & Yes                        & Single account               & Yes                                     & No                                                              & Yes                                                           & No human-readable TimeMaps   & No                          & Type 2  \\\midrule
HAW                          & Archived copies                              & Yes                       & Greater web archive projects & No                                      & Yes                                                             & No                                                            & Greater web archive         & No                   & Type 1         \\\midrule
Internet Archive (IA) user account web archives                     & Captures      & No & Single account & No & Yes & No & Greater web archive & No  & Type 2 \\
\midrule
LC                          & Captures                            & No                        & Greater web archive projects & No                                      & Yes                                                             & No                                                            & Greater web archive         & Yes            & Type 1               \\
\midrule
UKWA                         & Captures                            & Yes              & Greater web archive projects & No                                      & Yes                                                             & No                                                            & Greater web archive         & Yes                           & Type 2               \\ 
\midrule
PANDORA Subject/ Collection           & Webpage snapshots                   & Yes                    & Organizational collaborators & No                                      & Yes                                                             & No                                                            & Greater web archive         & No             & Type 1               \\
\midrule
Trove Collection             & Webpage snapshots                   & Yes                       & Organizational collaborators & No                                      & Yes                                                             & No                                                            & Greater web archive         & No                    & Type 2        \\ \bottomrule
\end{tabular}
}
* -- the \texttt{/all/} collection is an exception containing all mementos on Archive-It\\
\end{table*}

\subsection{Different Web Archive Platform Collection Structures' Features}

For each platform, we highlighted the similarities and differences in collection structures between platforms.

\begin{itemize}
\item \textbf{The term used to define mementos} 
\tabto{0.5cm} 
We see different names for mementos. The term \emph{memento} was formally established in RFC 7089 \cite{van_de_sompel_rfc_2013} in 2013. Many platforms predate that formality, thus different terms used exist for mementos. Some call them mementos, some call them copies, captures, or snapshots. Developers will need to understand the nomenclature of the platform and how these are synonymous. 

\item \textbf{Existence of sub-collections} 
\tabto{0.5cm} 
Most platforms support sub-collections, allowing an archivist to further narrow a collection's topic. These sub-collections have a variety of names, such as sub-collection (Trove, PANDORA), subject (PANDORA), or subcategory (HAW, PANDORA). Tools that encounter sub-collections must handle this hierarchy. 

\item \textbf{Attribution}
\tabto{0.5cm} 
Some web archives attribute curation to a single entity while others cite different organizational collaborators. UKWA, LC, and HAW all attribute the selection of their mementos to the greater web archive, meaning that the projects of the archiving organization as a whole led to the creation of these mementos. In contrast, PANDORA and Trove collections attribute memento selection to one or more organizations who requested their capture. Archive-It and Conifer, however, support individual accounts, thus memento selection is done by the person or organization that maintains that account. 

\item \textbf{Support for private collections} 
\tabto{0.5cm} 
Some web archives have put in place measures to control access so that other users of the web archive can or cannot view specific content. As account-based services, both Conifer and Archive-It allow private collections for users who are not yet ready to share their mementos. When a user makes a collection private, it means that the particular collection including all of its seeds will be private and can be only viewed and accessed while logged into your account.

\item \textbf{If the mementos (seed and deep) are accessible from more than one collection?} 
\tabto{0.5cm} 
On platforms like Archive-It and Conifer, each collection is isolated from one another. Archive-It supports an \texttt{/all/} collection containing mementos from every collection, but it is not advertised through the user interface. With this exception, a memento from one Archive-It collection is not accessible from another Archive-It collection. If two archivists create two Archive-It mementos from the same original resource at the same memento-datetime, but within different collections, then the mementos are distinct. This holds true for Conifer as well. Likewise, any deep mementos created from the crawls in these collections are only accessible to a user or tool browsing within the collection.

\item \textbf{Human-readable TimeMap membership} 
\tabto{0.5cm} 

Typically, a Human-readable TimeMap for an original resource will include all of the mementos of that URI-R in the web archive, rather than mementos within a collection. But Archive-It takes this a step further, with human-readable TimeMaps that only apply to an original resource as captured within the collection.

\item \textbf{If the resources are embargoed?} 
\tabto{0.5cm} 
Some archives limit access to their content offsite (locations other than their library premises). This is known as embargoing resources. Both LC and UKWA embargo resources. In both cases, some mementos are only available to patrons who physically visit their library campus. This creates challenges for Internet-based collection analysis tools because some mementos are hidden.
\end{itemize}

\subsection{Navigational Hierarchies}

\begin{figure*}
\centering
    \begin{subfigure}{0.5\hsize}
    \centering
    \includegraphics[width=0.8\textwidth]{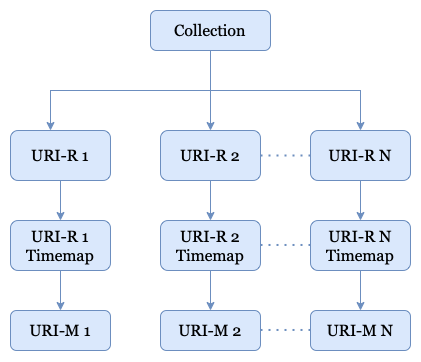}
    \caption{Navigational hierarchy: Type 1}
    \label{fig:type1-navigational-hierarchy}
    \end{subfigure}%
    ~
    \begin{subfigure}{0.5\hsize}
    \includegraphics[width=0.8\textwidth]{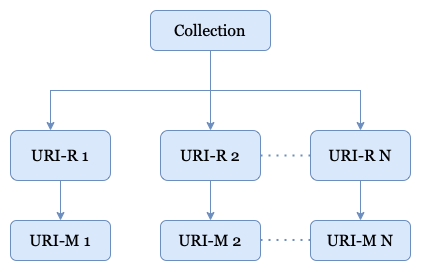}
    \caption{Navigational hierarchy: Type 2}
    \label{fig:type2-navigational-hierarchy}
    \end{subfigure}
\caption{The two main types of navigational hierarchies of different web archive collection platforms.}
\end{figure*}

In addition to the features of each web archive's collection structures, we also note different \textbf{navigational hierarchies}. These navigational hierarchies help us understand how a visitor or crawler navigate each collection for information. We identified two main navigational hierarchy types followed by the web archive platforms. Type 1 (Figure \ref{fig:type1-navigational-hierarchy}) allows the visitor to review a TimeMap before choosing a memento from the collection, thus an original resource supports the collection's theme. Type 2 (Figure \ref{fig:type2-navigational-hierarchy}) gives visitors direct access to mementos in a collection without having to go through a TimeMap, thus the memento supports the collection's theme.

\subsection{Different Web Archive Platforms}

\subsubsection{Archive-It}
\hfill\\
Archive-It is an Internet Archive subscription service where users can create their own collections. Figure \ref{fig:archiveit-navigational-hierarchy} illustrates the navigational hierarchy of an Archive-It collection. Each collection advertises a set of original resources (seeds). Each seed has its own TimeMap that provides mementos (seed mementos). Seed mementos link to deep mementos that are not shown on the collection landing page. The mementos (seed or deep) are always contained within the collection. This means that even if a visitor keeps browsing for more mementos (seed/deep) they will still stay with the initial collection in which they started. An Archive-It visitor following links from seed and deep mementos  never reaches a memento outside of the collection.

Within each collection, an original resource's mementos are listed in its TimeMap. Each TimeMap is specific to a collection and does not cross collections. Even though an original resource may appear in more than one collection, its mementos and its TimeMap do not. Archive-It collections do not support sub-collections. The archivist has the authority to add or remove seeds into the collection. The archivist schedules the crawls that create mementos. The archivist must, at a minimum, give their collection a name and supply seed URLs. From a visitor's perspective, the minimum amount of metadata provided by an Archive-It collection is limited to: the collection name, the collecting organization, the creation date of the collection, and the seed URLs. An archivist can, at their discretion, add more metadata to the collection or to individual seeds by choosing fields from Dublin Core \cite{dublin-core} or creating their own fields.

In the Archive-It service, the mementos are referred to as “captures”. The archivist has the option to make a collection publicly available to everyone or they can make it private, preventing others outside of their organization from viewing it.

Example URLs for Archive-It collection objects:

\begin{itemize}
\item Collection URL: \url{https://archive-it.org/collections/1064}
\item Seed URI: \url{http://beta.worldbank.org/climatechange/}
\item Human-readable TimeMap:  \url{https://wayback.archive-it.org/1064/*/http://beta.worldbank.org/climatechange/}
\item Machine-readable TimeMap (not seen by the user, but accessible to Memento clients): \url{https://wayback.archive-it.org/1064/timemap/link/http://www.worldbank.org/en/topic/climatechange}
\item Memento (seed memento): 
\url{https://wayback.archive-it.org/1064/20171206011529/http://www.worldbank.org/en/topic/climatechange}
\item Memento (deep memento) :
\url{https://wayback.archive-it.org/1064/20171206031816/http://www.worldbank.org/en/topic/climatechange/overview}
\end{itemize}

\subsubsection{Conifer}
\hfill\\
Conifer (formerly known as Webrecorder) is a service that allows a user to record and replay websites. With Conifer's navigational hierarchy, shown in Figure \ref{fig:conifer-navigational-hierarchy}, the visitor views the collection landing page. From this landing page, they can visit archivist-created lists \emph{lists} that serve as organized sub-collections. Inside each list is a set of page titles and URI-Rs, but these titles link directly to the memento that Conifer captured. From there a visitor can follow links to deep mementos. This service is different from other public archives that we have discussed because a Conifer user can use their browser to control the archiving process. Conifer allows users to create their own accounts with which they create and share their own collections. Each collection can be either public or private. Conifer has no concept of human-readable TimeMaps. Instead, mementos are grouped into ``sessions'' and with sessions, a user record web pages while browsing naturally. The account owner chooses the resources to preserve.

Example URLs for Conifer collection objects:

\begin{itemize}
\item Collections by a particular user: \url{https://conifer.rhizome.org/shawnmjones}
\item Public collection:  \url{https://conifer.rhizome.org/shawnmjones/wac_collection1}
\item Private collection: \url{https://conifer.rhizome.org/shawnmjones/wac_collection2}
\end{itemize} 

\subsubsection{Croatian Web Archive (HAW)}
\hfill\\
The Croatian Web Archive -- or Hrvatski Arhiv Weba (HAW) -- was built by the National and University Library in Zagreb in collaboration with the University of Zagreb University Computing Centre (Srce). Figure \ref{fig:haw-navigational-hierarchy} shows the navigational hierarchy of a HAW collection. The top HAW landing page lists the subjects. Clicking on a subject takes the visitor to a list of subcategories. Each subcategory/collection advertises a set of original resource titles and URI-Rs with a list of mementos. A visitor clicks on one of these URI-Rs to reach a human-readable TimeMap in list format. At HAW, the common term for mementos is ``copies'' and they are derived from the general web archive. The collections at HAW are publicly applicable to the users. These collections are compiled by the greater web archive.

Example URLs for HAW collections:
\begin{itemize}
\item HAW category/subjects: \url{https://haw.nsk.hr/en/category/biology-botany-and-zoology}
\item HAW subcategory/collection: \url{https://haw.nsk.hr/en/category/biology}
\item TimeMap: \url{https://haw.nsk.hr/en/publikacija/4818/}
\end{itemize}

\subsubsection{Internet Archive's (IA) User Account Web Archives}
\hfill\\
The Internet Archive is the largest and oldest web archive. It has its own types of collections that contain web archive files/data of different media types, but they do not present individual mementos for user consumption. However, there are collections that falls under the scope of our study, which we refer to as “IA’s user account web archives” which is a collection made by a particular user who has an account created at the Internet Archive. Its collections are tied to a specific user rather than a theme.

The navigational hierarchy of IA's user account web archives is shown in Figure \ref{fig:ia-navigational-hierarchy}. Once a visitor has reached a IA’s user account web archives collection, they click the name of a URI-R and directly get to the specific memento chosen for that collection. The design of these collections emphasizes that specific mementos, not all mementos for an original resource, are chosen as collection members. From each memento, a visitor can reach a human-readable TimeMap, allowing them to view other mementos for that same original resource.

Example URLs for Internet Archive's (IA) user account web archive collections:
\begin{itemize}
\item IA's user account web archive URL:  \url{https://archive.org/details/@shawnmjones?tab=web-archive}
\end{itemize}

\subsubsection{Library of Congress (LC)}
\hfill\\
At the Library of Congress, a visitor can access the available collections by clicking on the ``Digital Collections'' tab on the library home page. There are no sub-collections. The collections can contain other digital material (eg: images, video, PDF, etc.) besides web pages. For web pages, the Figure \ref{fig:loc-navigational-hierarchy} shows the navigational hierarchy of a LC collection. Once a visitor has selected a collection item, clicking on ``view captures'' on the image preview or caption will let the visitor access the TimeMap used to access the mementos. The collections are derived from the by the greater web archive itself. Mementos are referred to as ``captures''. The mementos are derived from the general web archive. The metadata to describe the collection contains, ``name'', ``description'', ``collection period'', ``frequency of collection'', ``languages'', and ``acquisition information.'' All collections at the Library of Congress are public, however, there are collection items that may contain embargoed content. Embargoed content is not accessible to those users who are off library premises or before a particular expiration date.

Example URLs for LC collections:
\begin{itemize}
\item Collection URL: \url{https://www.loc.gov/collections/egyptian-elections-web-archive/}
\item Collection Item (full access only at the library):  \url{https://www.loc.gov/item/lcwaN0006607/}
\item Collection Item (content may be embargoed):  \url{https://www.loc.gov/item/lcwaN0006608/}
\item Collection Item (images + web pages):
\url{https://www.loc.gov/collections/tenth-to-sixteenth-century-liturgical-chants/}
\end{itemize}

\subsubsection{The United Kingdom Web Archive (UKWA)}
\hfill\\
At the UKWA, the collections are listed as  ``Topics and Themes'' on the home page. The navigational hierarchy of a UKWA collection is shown in Figure \ref{fig:ukwa-navigational-hierarchy}. Once a visitor to the UKWA has reached a collection or sub-collection, they click the name of a URI-R and directly reach a memento specifically chosen for that collection similar to the IA's user account web archives. The UKWA curates the collections themselves. Each collection consists of a set of individual mementos. Once a visitor selects a collection, they can search (using URI-R or keyword) within the collection. In addition to mementos, a collection may contain sub-collections. Mementos come from the general UKWA; they are not bound to a collection. From each memento, a visitor can reach a human-readable TimeMap, allowing them to view other mementos for that same original resource. In the UKWA, mementos are referred to as ``captures''. Although all collections are public, most of mementos are only viewable from the library premises. 

Example URLs for UKWA collections:

\begin{itemize}
\item Collection URL (without sub-collections): \url{https://www.webarchive.org.uk/en/ukwa/collection/44}
\item Collection URL (with sub-collections):
\url{https://www.webarchive.org.uk/en/ukwa/collection/910}
\item Sub-collection URL: \url{https://www.webarchive.org.uk/en/ukwa/collection/911}
\item Memento (available outside library):
\url{https://www.webarchive.org.uk/wayback/archive/20161128182225/http://www.iwa.wales/click/2016/07/referendum-ukips-role-now/}
\item Memento (available inside library only) are linked to: \url{https://www.webarchive.org.uk/en/ukwa/noresults}
\end{itemize}

\subsubsection{National Library of Australia (NLA)}
\begin{figure*}
\centering
\includegraphics[width=\textwidth]{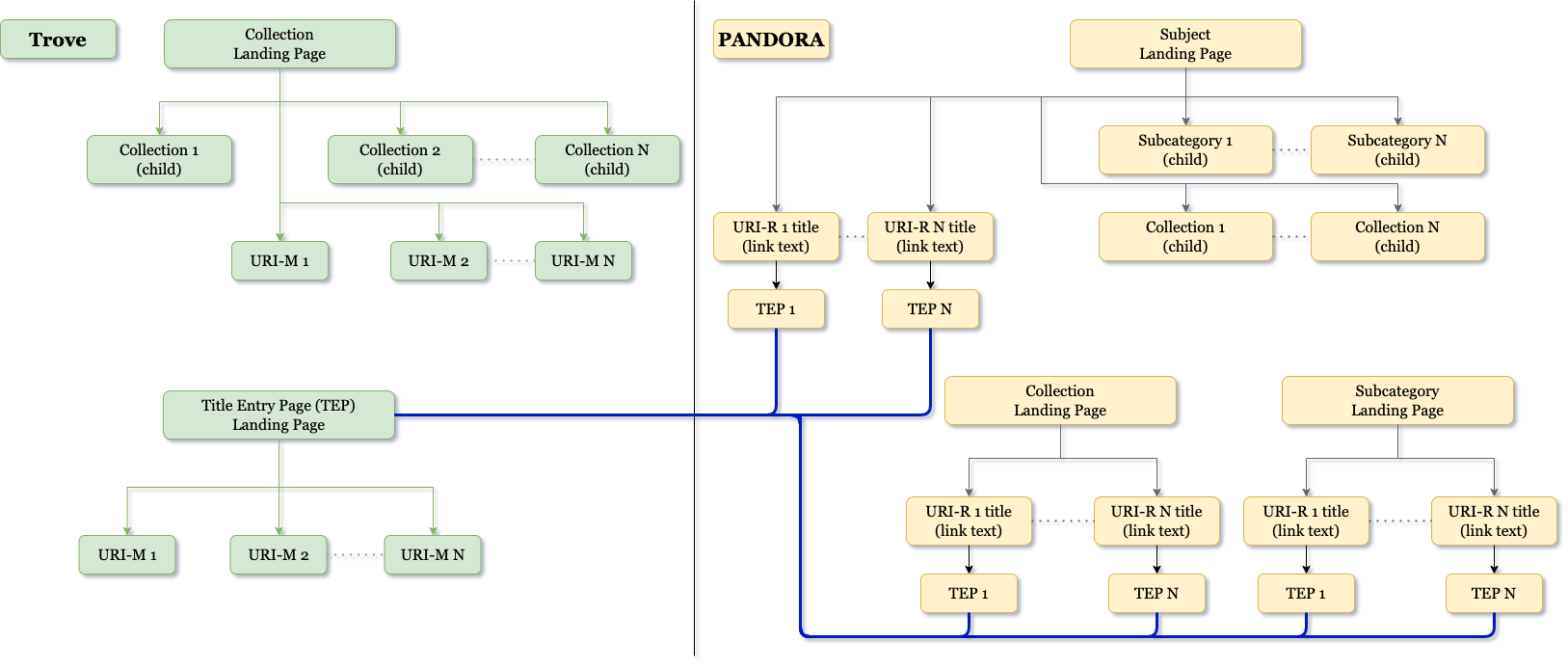}
\caption{The navigational hierarchies of the collections at the NLA.}
\label{fig:nla-navigational-hierarchy}
\end{figure*}

\hfill\\
Trove \footnote{https://trove.nla.gov.au/} is an initiative between the National Library of Australia (NLA) and other Australian partner institutions. The NLA and these partner organizations decide which resources to preserve. In the National Library of Australia, Trove is the new discovery system and PANDORA is the much older Australian web archive. Figure \ref{fig:nla-navigational-hierarchy} shows the complex navigational hierarchy of PANDORA and Trove. This navigational hierarchy exists because NLA has been migrating its mementos to Trove but still wishes to retain the effort put into creating collections at PANDORA. Thus, a visitor has several potential points of entry. If the visitor comes from Trove, then their path is much like with UKWA and IA's user account web archives, going from collection directly to its mementos. All Trove mementos come from the greater web archive and are not restricted to a collection as with Archive-It. These mementos link to other deep mementos which are not featured on the collection page. Trove collections can contain mementos and sub-collections.

If the visitor comes from a PANDORA Subject or Collection, then they are presented with a list of titles representing original resources chosen for inclusion in the collection. Clicking on one of these titles takes the user to a Title Entry Page (TEP), which is a human-readable TimeMap. From the TEP, the visitor can then choose a memento hosted at Trove. This means that ultimately, PANDORA subjects and collections are linked to Trove TEPs making a link between Trove and Pandora. Additionally, each TEP is informed by data stored at a URL like  (\url{https://webarchive.nla.gov.au/bamboo-service/tep/{TEP_ID}}) that returns a JSON object with the URI-Ms and other metadata about the TEP. This JSON resource acts like a machine-readable TimeMap of its own kind. A TEP's human-readable metadata contains the original resource page title and a list of collecting organizations (as “partner organizations”) and their logos. In both Trove collections and TEPs the mementos are commonly referred to as “webpage snapshots”. All the collections at Trove are public.

The home page of PANDORA lists the main subjects under ``Browse subjects''. A PANDORA subject may contain “subcategories” that also fall under the PANDORA subject category and “collections”. In addition to TEP pages, some of these PANDORA collections can contain sub-collections. All Pandora subjects and collections are public. At Pandora (both Pandora subjects and collections), the mementos are referred to as ``webpage snapshots''. These mementos are derived from the greater web archive at Trove; they are not bound to a collection as with Archive-It. 

Example URLs for NLA collections:

\begin{itemize}
\item Trove collection: \url{https://webarchive.nla.gov.au/collection/15003} 
\item Trove sub-collection: \url{https://webarchive.nla.gov.au/collection/15052}
\item The Pandora subject URL for “Arts”: \url{http://pandora.nla.gov.au/subject/2}
\item A Pandora subcategory URL of “Arts” named “Dance”: \url{http://pandora.nla.gov.au/subject/42}  
\item A Pandora collection URL: \url{https://pandora.nla.gov.au/col/12142}
\item A Pandora sub-collection URL: \url{https://pandora.nla.gov.au/col/12203}
\item A Trove TEP URL: \url{https://webarchive.nla.gov.au/tep/88147}
\end{itemize}


\begin{figure*}
\centering
    \begin{subfigure}{0.5\hsize}
    \centering
    \includegraphics[width=\textwidth]{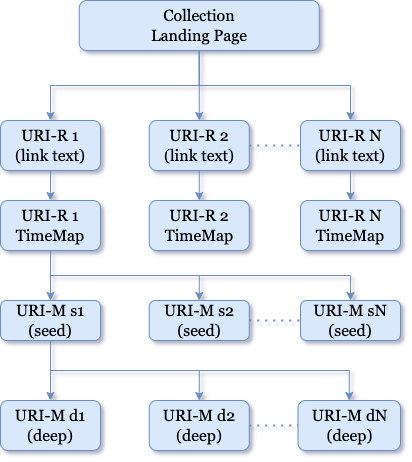}
    \caption{Archive-It}
    \label{fig:archiveit-navigational-hierarchy}
    \end{subfigure}%
    
    \begin{subfigure}{0.5\hsize}
    \includegraphics[width=\textwidth]{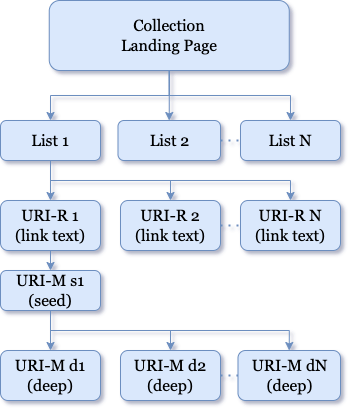}
    \caption{Conifer}
    \label{fig:conifer-navigational-hierarchy}
    \end{subfigure}
\caption{The navigational hierarchies of different web archive collection platforms.}
\end{figure*}

\begin{figure*}\ContinuedFloat
\centering
    \begin{subfigure}{0.5\hsize}
    \centering
    \includegraphics[width=\textwidth]{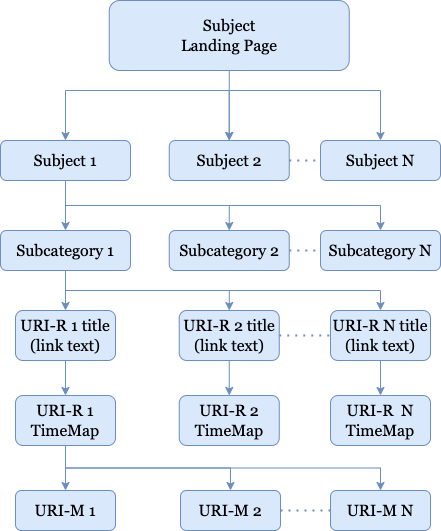}
    \caption{Croatian Web Archive (HAW)}
    \label{fig:haw-navigational-hierarchy}
    \end{subfigure}%

    \begin{subfigure}{0.5\hsize}
    \includegraphics[width=\textwidth]{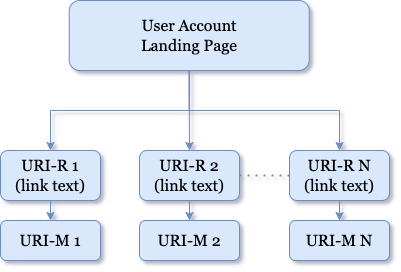}
    \caption{Internet Archive (IA) user account web archives}
    \label{fig:ia-navigational-hierarchy}
    \end{subfigure}

\caption{The navigational hierarchies of different web archive collection platforms.}
\end{figure*}

\begin{figure*}\ContinuedFloat
\centering
    \begin{subfigure}{0.5\hsize}
    \includegraphics[width=\textwidth]{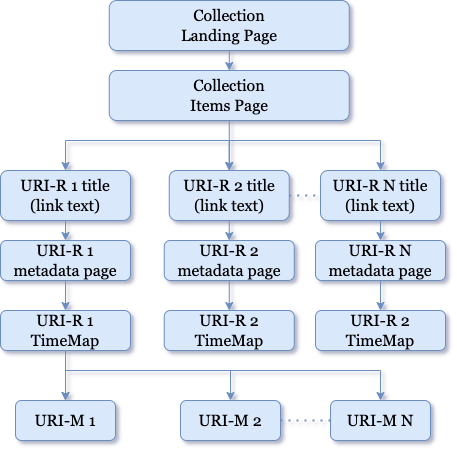}
    \caption{Library of Congress (LC)}
    \label{fig:loc-navigational-hierarchy}
    \end{subfigure}

    \begin{subfigure}{0.5\hsize}
    \includegraphics[width=\textwidth]{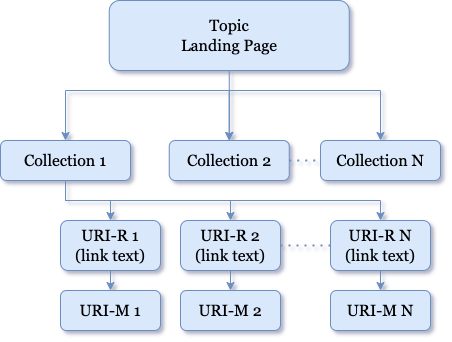}
    \caption{United Kingdom Web Archive (UKWA)}
    \label{fig:ukwa-navigational-hierarchy}
    \end{subfigure}%
\caption{The navigational hierarchies of different web archive collection platforms.}
\end{figure*}

\section{Future Work}
One could extend our research to understand the any socioeconomic, political, or management factors that influence how each collection structure is arranged.
We also intend to study collection structures of other national web archives such as the Portuguese Web Archive \footnote{https://arquivo.pt/} and the Icelandic Web Archive \footnote{https://vefsafn.is/}. We would also like to understand the structure and features of web archiving initiatives run by universities, such as Stanford University Libraries \footnote{https://library.stanford.edu/spc/university-archives} and Columbia University Libraries \footnote{https://library.columbia.edu/index.html}. We would like to understand why some web archives do not yet have collections. Is it related to the size of the web archive? Perhaps there are social, cultural, or management challenges that are not visible from examining the web archive itself? Finally, with this understanding we intend to suggest enhancements and improvements to tools like AIU using this knowledge of collection structures.

\section{Conclusion}

Web archives play a role in preserving our digital history. As web archives grow, archivists eventually create collections to make their archives easier to understand and manage. Collections help visitors narrow the number of documents they need to review for a specific topic. Web archive collections are also targets for different data management and analysis tools. Each collection's structure influences how it meets these different use cases.


We have addressed RQ1 by reviewing the the collection structures of Archive-It, NLA (Trove and PANDORA), HAW, LC, UKWA, and Conifer. Through this analysis we discovered different approaches to collection structure, revealing a diversity of capabilities for potential new web archiving platforms and tools. These approaches appear to be informed heavily by the nature of each platform. For RQ2 we sought similarities in these collection structures so we could better understand their current state. Archive-It and Conifer are account-centric, meaning that an individual user or organization cares for the collection separately from the archiving platform itself. The others are general web archives that created collections from their vast holdings based on internal projects. Where mementos Archive-It or Conifer's mementos are accessible within a collection, the other platforms share mementos between collections. Archive-It or Conifer offer direct curatorial attribution to the account owner, Trove and PANDORA cite different organizational collaborators who requested the creation of mementos, and the others are inconsistent in this regard. Most archives offer at least one level of sub-collection and some web archives embargo resources.

We discovered two types of navigational hierarchies for collections. In Type 1, an original resource supports the collection’s theme. In Type 2, a memento supports the collection’s theme. In Type 1 navigational hierarchy, used by platforms like Archive-It and LC, the user navigates through zero or more sub-collections before reaching a human-readable TimeMap. From there, they can select the memento of their choice. In Type 2 navigational hierarchy, used in plaforms like UKWA and Trove, the user navigates through zero or more sub-collections before reaching a page that directly links to mementos. Visitors and tool designers need to understand this distinction. These different structures reflect different decisions on resource membership among collections.

We did not attempt to prescribe how a web archive collection should be designed but rather analyzed existing platforms. Such information is helpful to different parties. Future archivists and platform designers need ideas for their own archives. Software developers need to understand how to process these collection structures to build tools. With the growth in interest in Big Data, web archives will increasingly become the target of interest for researchers. With an understanding of collection structures, researchers will know how to acquire the metadata and mementos they seek.

\bibliographystyle{ACM-Reference-Format}

\newpage
\bibliography{refs}

\end{document}